\documentclass[prl,aps,
nofootinbib,  
superscriptaddress,showpacs,twocolumn,longbibliography]{revtex4-1}

\usepackage{graphicx} 
\usepackage[colorlinks]{hyperref}
\usepackage{amsmath,amssymb, physics} 
\usepackage[dvipsnames]{xcolor}

\usepackage{bm} 

\usepackage{bbold}  

\newcommand{\beq}{\begin{equation}}
\newcommand{\eeq}{\end{equation}}

\usepackage{mathtools}
\usepackage{comment}
\usepackage{amsmath} 

\hypersetup{linkcolor=magenta,
    citecolor=blue}

\begin{document}

\title{{Supermartingales in Quantum Resources Theories:\\
Where do quantum resources go when you're watching?}}

\author{Katarzyna Macieszczak}
\affiliation{Department of Physics, University of Warwick, Coventry CV4 7AL, United Kingdom}
\author{Robert L. Jack}
\affiliation{Department of Applied Mathematics and Theoretical Physics, University of Cambridge, Wilberforce Road, Cambridge CB3 0WA, United Kingdom}
\affiliation{Yusuf Hamied Department of Chemistry, University of Cambridge,
Lensfield Road, Cambridge CB2 1EW, United Kingdom}


\begin{abstract}
We establish a connection between quantum resource theory and probability theory, under repeated application of free operations. We show that strong monotonicity of a resource measure implies that the resource is a supermartingale.  We then use the optional stopping and martingale convergence theorems to derive bounds on the efficiency of post-selection, and other free adaptive strategies. We {also} describe the asymptotic dynamics of the conditional state, where resource fluctuations are necessarily absent, showing one of two distinct phenomena occurs: resource vanishing or resource freezing.
\end{abstract}

\maketitle

Quantum technologies promise ground-breaking improvements in computing, cryptography, and many other areas~\cite{Bouwmeester2010,Nielsen_Chuang_2010}. 
\emph{Quantum resource theories} (QRTs)~\cite{chitambar2019rmp} provide valuable insight into the feasibility and success rates of new devices and algorithms.  For example, {quantum} entanglement~\cite{Schroedinger1935,Horodecki2009} can be characterised as a resource {for} quantum cryptography and communication {and  cannot be created by local operations and classical communication (LOCC)}.  It is useful {generally} to identify a class of free operations that do not introduce any resource, or {more practically,} may be feasibly performed in particular experimental setups.  The amount of a resource {in a quantum state} can be quantified by a resource measure; {those often carry operational interpretation so that}  states with more resource are more powerful in {relevant} applications.

For quantum systems that are subject to {(generalised)} measurements, 
the randomness of measured outcomes leads to stochastic dynamics~\cite{Wiseman_Milburn_2009}, and the resource may become a random variable.  The additional information from the outcomes allows one to (indirectly) "watch"  the system ~\cite{Murch2013,Guerlin2007,Weber2014}, which can be exploited for greater control in experiments~\cite{Sayrin2011,Vijay2012,Rossi2018}. Measurements also
enable new phenomena such as purification~\cite{maassen2006,Benoist2019}, in which initial mixed states converge to pure states at large times.   In systems with strong symmetries~\cite{Buca2012,Albert2014},  coherences between symmetry subspaces tend to vanish at large times in quantum trajectories; this phenomenon has been termed dissipative freezing in~\cite{Munoz2019}.  Both purification and freezing are concerned with the long-time behaviour of quantum trajectories, which converge at large times to particular classes of quantum states~\cite{phdVanHorssen,Benoist2024,schmolke2025}.

This work shows how such phenomena can be understood generically, by establishing a relationship between  quantum resources {theories with so-called strongly-monotonic~\cite{Vidal2000} resource measures}, and the probabilistic theory of martingales~\cite{williamsBook,Liptser_2001,Roldan2023}.  
Then the celebrated \emph{{martingale} optional stopping theorem} leads to a general bound on the effectiveness of \emph{post-selection} (and other) methods for the distillation~\cite{Bennett1996a} ({stochastic} increase) of quantum resources.  
Moreover, the \emph{martingale convergence theorem} places strong constraints on the long-time behaviour of trajectories, in such cases.
For systems with strong symmetries, this reveals that
{the long-time behaviour in~\cite{Munoz2019}} corresponds to the loss of {quantum} coherence~\cite{vaccaro2008,baumgratz2014,streltsov2017} at large times, which is but one example of a  general phenomenon that we call \emph{resource vanishing}.  In fact, if the resource is a supermartingale then it must either vanish asymptotically, or converge to some constant positive value, which we call {resource freezing}.  
Earlier work on freezing quantum resources, such as quantum correlations~\cite{cianciaruso2015} or quantum coherence~\cite{bromley2015} found conditions where some non-zero resource is unaffected by certain free operations, no matter how the resource amount is quantified.  We show instead that quantum resources can be frozen \emph{forever} in stochastic quantum trajectories.

In addition to establishing the general framework outlined above, this work illustrates its implications with two examples, where the relevant resources are {quantum} entanglement and coherence.  Strong monotonicity~\cite{Vidal2000} is also considered in other QRTs, such as that of magic~\cite{Veitch_2012,Veitch_2014,Gu2025} relevant for stabiliser quantum computation, {and always holds for resource measures based on convex roof extensions~\cite{Wei2003,Uhlmann2010,Streltsov_2010}}.

\emph{Quantum channels and stochastic dynamics} --  A quantum channel $\Lambda$ is a completely positive (CP) and trace-preserving (TP) linear map that acts on a density matrix $\rho$ of a quantum system with a finite-dimensional Hilbert space $\mathcal{H}$. 
It describes non-unitary evolution, which might be due to coupling to an external environment, or (projective or generalised) measurements.   We consider {repeated} applications of $\Lambda$, that is
\beq
\rho_{t+1} = \Lambda(\rho_t), 
\label{eq:average-state-evol}
\eeq
where $\rho_t$ is the state after $t$ timesteps (with $\rho_0$ the initial state).
Note, the dynamics of Eq.~\eqref{eq:average-state-evol} is \emph{deterministic}.

Any channel $\Lambda$ can be decomposed into (a set of) \emph{subchannels} ${\cal S}_\Lambda=\{\Lambda_1,\Lambda_2,\dots,\Lambda_M\}$ as
\beq
\Lambda(\rho)  = \sum_{m=1}^M \Lambda_m(\rho),
\label{eq:channel_subchannel}
\eeq
such that every $\Lambda_m$ is CP, but only their sum is TP.
A \emph{stochastic} dynamics  is obtained by inferring (or measuring) which subchannel occurred in each time step.   This leads to a time-dependent (random) conditional state, denoted by $\varrho_t$.
The stochastic dynamics is that 
with probability 
\beq
p_m(\varrho_t) = \Tr[ \Lambda_m(\varrho_t)],
\eeq
the conditional state at step $t+1$ is   
\beq
\varrho_{t+1} = \frac{ \Lambda_m(\varrho_t) }{ p_m(\varrho_t) } 
\label{eq:stochastic-state-evol}
\eeq
for $m=1,...,M$.   The label $m$ might be the outcome of a projective measurement, or a more general inferred outcome within the theory of quantum-classical measurement maps (QCMMs), {which include so-called physically implementable operations~\cite{Chitambar2016,Marvian2016}}. In particular ${\cal S}_\Lambda$ defines a corresponding QCMM, see the Supplemental Material for further details.

\emph{Quantum Resource Theories} -- 
A resource in QRT is some property of (mixed) quantum states, such as their entanglement or their coherence in some relevant basis.  We give an outline of this {framework}~\cite{chitambar2019rmp}, 
with further details in the Supplemental Material~\cite{SM}.

 \emph{Free states} in QRT contain no resource. \emph{Free operations} cannot create resource, so they transform any free state into (another) free state.  The resource is quantified by a  \emph{resource measure} $R$; a non-negative function such that every free state $\rho$ has $R(\rho)=0$ and $R$ {does not increase} under free operations.  Hence, if the application of channel $\Lambda$ is a free operation then
\beq
R[\Lambda(\rho)] \leq R(\rho).
\label{eq:mono}
\eeq
for all states $\rho$.

 A given resource may be quantified by several different resource measures.  A measure is  \emph{faithful} if it vanishes {for} all free states. 
Resource measures are 
often \emph{continuous} functions of the system state; they may also be \emph{bounded}.
 These properties are not required for our main results but they are relevant in certain cases, see below.

\emph{Strong monotonicity} -- For a QRT with {a} resource measure $R$, we consider a quantum channel $\Lambda$ that corresponds to a free operation, and is decomposed into subchannels following Eq.~\eqref{eq:channel_subchannel}.  The resource measure may be non-monotonic along trajectories, but a central assumption of this work is that it does not increase on average, that is
\beq
\sum_{m=1}^M p_m(\rho) \,R\!\left[ \frac{\Lambda_m(\rho)}{p_m(\rho)} \right] \leq  R(\rho)  
\label{eq:strong-mono}
\eeq 
for all $\rho$.
If Eq.~\eqref{eq:strong-mono} holds, then we say that \emph{$R$ is strongly monotonic for the subchannels ${\cal S}_\Lambda$}.  
We explain in~\cite{SM} that 
strong monotonicity is a natural property of QRTs with free QCMMs.

\emph{Connection to supermartingales} -- 
 The evolution of Eq.~\eqref{eq:stochastic-state-evol} is a Markov chain for the conditional state $\varrho_t$. 
We assume throughout that the initial state is fixed, $\varrho_0=\rho_0$.
A special case of Eq.~\eqref{eq:channel_subchannel} is that $\Lambda_m(\rho)=K_m \rho K_m^\dag$ where each $K_m$ is a Kraus operator.  In this case  
the decomposition of $\Lambda$ is {known as} an \emph{unravelling} of the channel, {and} the sample paths of the stochastic dynamics are called \emph{quantum trajectories}~\cite{Wiseman_Milburn_2009}.  Here, we consider the more general case where each $\Lambda_m$ may involve more than one Kraus operator. 
Nevertheless, we use the language of quantum trajectories to discuss the stochastic dynamics.

From Eq.~\eqref{eq:stochastic-state-evol}, we identify the expectation of $\varrho_{t+1}$, conditional on the state at time $t$ as
\beq
\mathbb{E}[\varrho_{t+1}|\varrho_t] = \Lambda(\varrho_t). 
\eeq
That is, averaging the stochastic dynamics recovers the channel $\Lambda$.
Hence the expectation value of the conditional state obeys  
	\beq
	\mathbb{E}[\varrho_t] = \rho_t 
	\label{eq:rho-ave}
	\eeq
and the averaged state evolves according to Eq.~\eqref{eq:average-state-evol}.

\begin{figure}
\includegraphics[width=8.5cm]{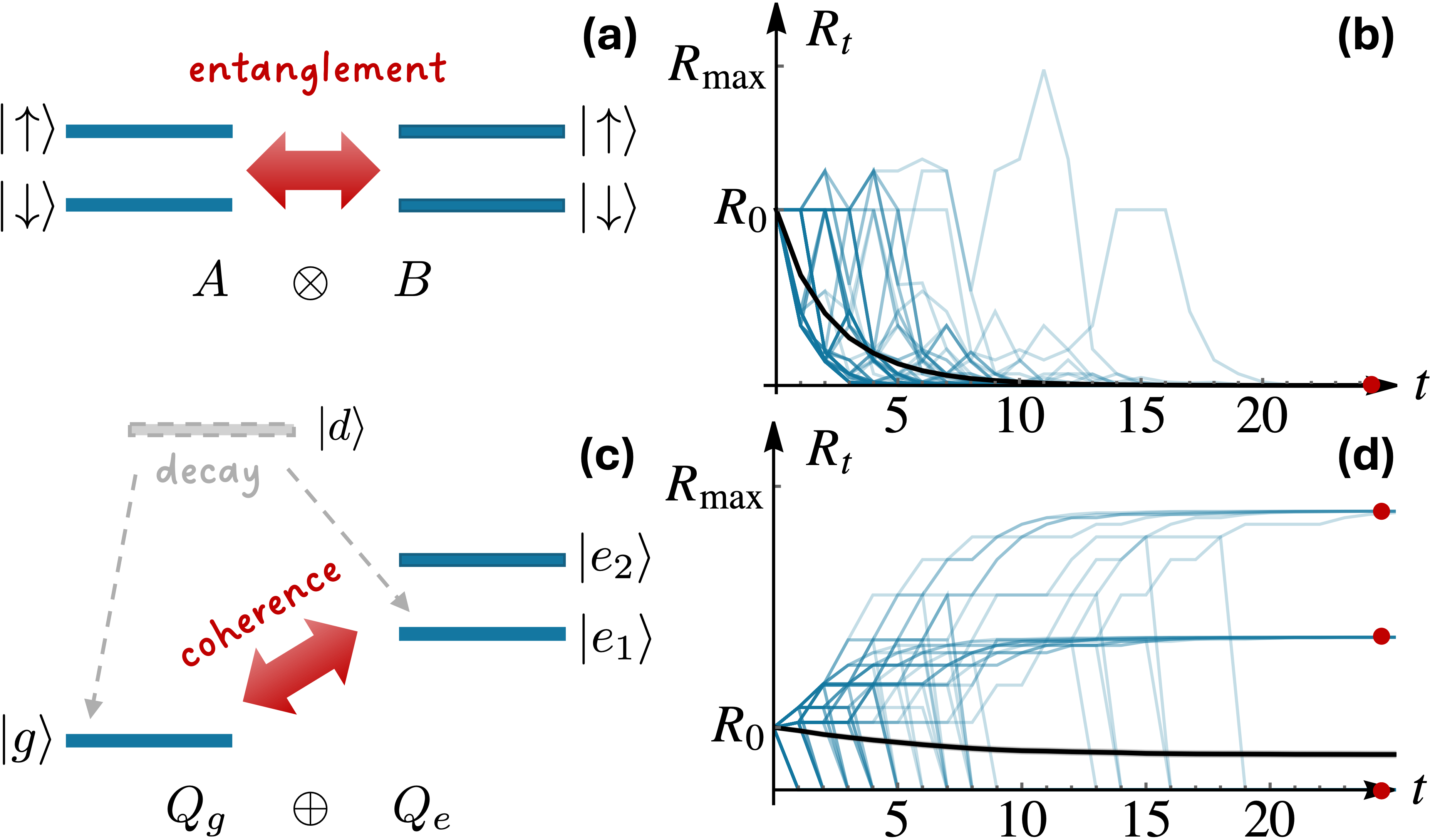}
\caption{\textbf{Illustrative {examples}.}  (a)~Example~I: two-qubit system where the resource is {quantum} entanglement. 
(b)~The resource measure $R_t$ along 100 representative quantum trajectories (blue) and the average resource (black).  (c)~Example~II: four-state system where the resource is {quantum} coherence between ground state and excited states.  (d)~The resource measure for representative trajectories.}
\label{fig:illustrate}
\end{figure}

We write $R_t=R(\varrho_t)$ for the resource measure of the conditional state, also $R_0=R(\rho_0)$.
The LHS of Eq.~\eqref{eq:strong-mono} is the expectation of the resource, on applying the subchannels to $\rho$.  For a strongly monotonic resource, this means
\beq
\mathbb{E}\big[ R_{t+1} \big| \varrho_t \big] \leq R_t \qquad \forall \, \varrho_t.
\label{eq:supermart}
\eeq
Hence, the average resource  is non-increasing:
\beq
\mathbb{E} [R_{t+1}]  \leq \mathbb{E} [R_t]. 
\label{eq:av_R}
\eeq
However, despite this averaged behaviour, it is crucial for this work that the resource $R_t$ itself has non-trivial fluctuations; in particular, it  may increase along individual quantum trajectories.  These fluctuations are crucial in what follows.  The general setup is illustrated in Fig.~\ref{fig:illustrate} by two specific examples, which are discussed below.

Our key observation is that Eq.~\eqref{eq:supermart} means that $R_t$ is a \emph{supermartingale}~\cite{williamsBook,Liptser_2001,Roldan2023}, see Appendix for further detail.  We now apply
two classical theorems from martingale theory in the context of QRT, from which we obtain new insights into the dynamics of the quantum resource and the conditional state.

\newcommand{\Tmax}{T}

\emph{Post-selection and optional stopping} -- Post-selection in QRT enables preparation of quantum states with resource measure above a threshold $R^*> R_0$: One generates quantum trajectories and follows the conditional state until $R_t \geq R^*$, the first time that this occurs is the stopping time ${\tau}$.  
However, if the time reaches a cutoff $\Tmax$ before the value $R^*$ is achieved then one abandons the post-selection attempt, and  sets {$\tau=\Tmax$}.  Since $R_t$ is a supermartingale and the stopping time {$\tau$} is bounded, one may apply the \emph{martingale optional stopping theorem}~\cite{williamsBook,Roldan2023}, which yields a generic constraint on post-selection (see Appendix):
\beq
\mathbb{E}[ R_{{\tau}} ] \leq R_0 \; ,
\label{eq:stopping}
\eeq
where $R_\tau$ is the value of $R_t$ at the stopping time.
Note, the expectation includes both the successful post-selections and the abandoned attempts.
This shows that post-selection cannot increase a resource on average -- any increase in the post-selected trajectories must be compensated by abandoned attempts where the resource is lower.  
The theorem further implies (see Appendix) that $\mathbb{E}[R_\tau]$ must decrease with $R^*$, because lower thresholds are reached earlier.

The probability of a successful post-selection is 
{$p^*=\mathbb{P}[R_{\tau}\geq R^*]$, which depends on general on the cutoff $T$.
Since the resource always exceeds $R^*$ in successful outcomes, ${p^*} R^* \leq \mathbb{E}[ R_{{\tau}} ]$. 
Then Eq.~\eqref{eq:stopping} gives a limit on the success probability:
\beq
{p^*} \leq \frac{ R_0 }{ R^* } \;   ,  
\label{eq:prob}
\eeq
which is again decreasing with $R^*$.  The calculation relies solely on the structure of the QRT, independent of any details of the specific channels, {cutoff time}, or resources. 

 Eq.~\eqref{eq:stopping} relies on a finite cutoff $T$, but   Eq.~(\ref{eq:prob}) also holds for bounded resource measures $R$ if there is no cutoff (see Appendix).
Post-selection can also be modelled as a QCMM, in which case ${p^*}$ 
is the sum of single-shot {stochastic} convertibility rates~\cite{Bennett1996a,Vidal2000} for all states from the QCMM with $R\geq R^*$, see~\cite{SM}.
Moreover, Eq.~\eqref{eq:stopping} remains valid, even if a different set of subchannels is chosen adaptively in each step, as long as $R$ remains strongly monotonic.  This statement is formalised in the Appendix, it means that the result in Eq.~\eqref{eq:stopping} cannot be avoided by using adaptive channels. This is our first general result: \emph{no resource can be increased on average by applying subchannels for which $R$ is strongly monotonic}.
The bound Eq.~\eqref{eq:prob} is similarly general.

\emph{Resource vanishing or resource freezing, and convergence} -- Our second general result exploits the \emph{martingale convergence theorem}~\cite{williamsBook,Liptser_2001,Roldan2023}.  This states that $R_t$ stops  fluctuating at large times, as seen in the examples of Fig.~\ref{fig:illustrate} (see also the analysis of purification in~\cite{maassen2006}).
It converges to an asymptotic value $R_\infty$.  Technically, this is almost-sure convergence, which means the limit
\beq
R_\infty = \lim_{t\to\infty} R_t  
\label{eq:conv}
\eeq
 exists with probability {1} (see Appendix).  
Since $R_t$ is still a supermartingale for adaptive channels, this result also rules out oscillations of $R_t$ in periodic (Floquet) dynamics where one applies a repeated sequence of strongly-monotonic channels.

In the example of Fig.~\ref{fig:illustrate}(b), $R_\infty=0$ for all trajectories, that is $\mathbb{P}[R_\infty=0]=1$.  We call this \emph{resource vanishing}.  
In Fig.~\ref{fig:illustrate}(d), the limit $R_\infty$ can be non-zero; moreover,
 $R_\infty$ is a random variable, taking different values for different trajectories.  If $\mathbb{P}[R_\infty>0]>0$,  we call this \emph{resource freezing}.
Our assumption of strong monotonicity means that either $R$ freezes, or it vanishes.  
Recalling that the same resource might be quantified by two different strongly-monotonic measures $R,\tilde{R}$, we show in the Appendix that if $R,\tilde{R}$ are both continuous and faithful, then either they both vanish, or they both freeze.  In this sense, the distinction between freezing and vanishing is a feature of the QRT itself, not the specific resource measure.

 \begin{figure}
\includegraphics[width=8.5cm]{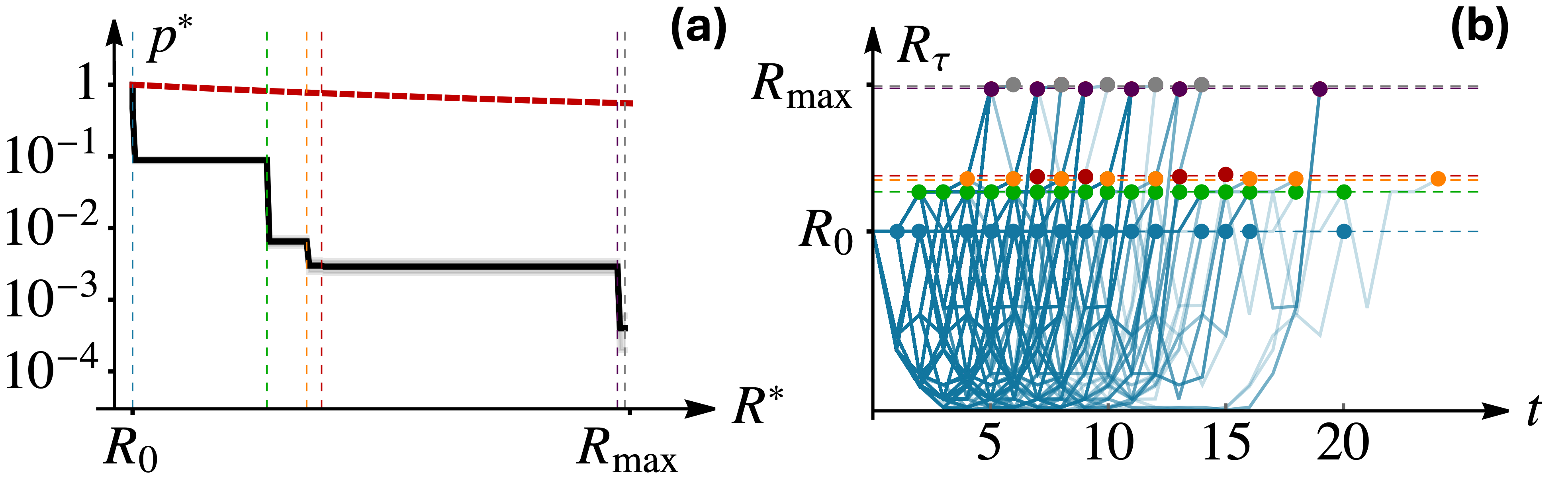}
\caption{\textbf{{Entanglement} post-selection in Example I}.  
(a)~Probability $p^*$ of successful post-selection (black) with $T=25$ and various thresholds $R^*$, compared with the bound of Eq.~\eqref{eq:prob} (red).
(b)~Stopped trajectories (blue) from (a) for selected values of $R^*$ (dashed lines); markers indicate final values. 
}
\label{fig:post_selection}
\end{figure}

 It is important that while $R_t$ converges to a limit, the conditional state $\varrho_t$ may continue to fluctuate at large times.
In this case, Eq.~\eqref{eq:conv} constrains the large-time behaviour of $\varrho_t$, which approaches an isosurface of $R$ [the set of states with $R(\varrho)=R_\infty$].  Then
the {quantum} trajectory is {asymptotically supported} on this isosurface (see Appendix for details).  The resource measure $R$ can often be interpreted geometrically as a distance to the nearest free state~\cite{chitambar2019rmp}, in which case isosurfaces are equally distant from the set of free states, {and quantum trajectories inherit this property, asymptotically}.

\newcommand{\upket}{|\!\uparrow\rangle}
\newcommand{\downket}{|\!\downarrow\rangle}
\newcommand{\upbra}{\langle\uparrow\!|}
\newcommand{\downbra}{\langle\downarrow\!|}

\emph{Example I : {entanglement post-selection}}.  Fig.~\ref{fig:illustrate}(a) illustrates a system with two qubits $A,B$ where the resource is the {quantum} entanglement~\cite{Horodecki2009}, so the set of free operations are local operations (LOs) with classical communication (CC).  {The resource measure $R$ is the entanglement of formation~\cite{Bennett1996b,Wootters1998}}.
The (free) two-qubit channel $\Lambda$ is composed of three steps: (i) a LO (one-qubit channel) $\Lambda^{\rm local}$ on qubit $A$ that {is a generalised measurement with two possible outcomes}; (ii) CC of the outcome to qubit $B$; (iii) depending on the communicated outcome, one either applies  $\Lambda^{\rm local}$ to $B$, or a unitary LO.  The outcomes of the measurements determine  three subchannels.
The resource measure $R$ is the entanglement of formation which is strongly monotonic for {LOCC} subchannels.  The system is initialised in a  pure state $\rho_0=|\psi\rangle\!\langle\psi|$ with $|\psi\rangle=(|\!\uparrow\uparrow\rangle + |\!\uparrow\downarrow\rangle +|\!\downarrow\downarrow\rangle)/\sqrt{3}$.
 It is crucial that $\Lambda^{\rm local}$ is not unitary, so the resource varies along individual quantum trajectories. Full details are given in SM~\cite{SM}.

This setup enables post-selection, for which results are shown in Fig.~\ref{fig:post_selection}.   We recall from Fig.~\ref{fig:illustrate}(b) that the average resource decreases with time, but Fig.~\ref{fig:post_selection}(b) shows how various values of $R^*$ can be achieved for stopped {quantum} trajectories.  The success probability $p^*$ decreases with $R^*$; it is bounded above by Eq.~\eqref{eq:prob}.  The bound is loose because the inequality in Eq.~\eqref{eq:strong-mono} is also loose for this channel; hence the average resource decays quite rapidly, which limits the post-selection efficiency (see the Appendix for further details).

\emph{Example II : {coherence freezing}} -- Fig.~\ref{fig:illustrate}(c) illustrates a four-level system where the quantum resource is the {quantum} coherence~\cite{vaccaro2008,baumgratz2014,streltsov2017} between the ground state $|g\rangle$ and two excited states $|e_1\rangle,|e_2\rangle$; the fourth state $|d\rangle$ is a decay subspace.  The resource measure $R$ is the relative entropy of coherence~\cite{vaccaro2008,baumgratz2014}.  We  take subchannels $\Lambda_m(\cdot)= K_m(\cdot)K_m^\dag$  with
\begin{align}
    K_1 & = |+\rangle\!\langle e_1| +
                  |g\rangle\!\langle g|/\sqrt{2}  +
                  |d\rangle\!\langle d|/\sqrt{2}  
    \;,
    \nonumber\\
    K_2 & = |-\rangle\!\langle e_2| +
                   |g\rangle\!\langle g|/\sqrt{2}  +
                  |d\rangle\!\langle d|/2  
    \;,
    \nonumber\\    
    K_3 & = |e_1\rangle\!\langle d|/\sqrt{8} \;, \qquad K_4 = |g\rangle\!\langle d|/\sqrt{8} \;,
\label{eq:kraus-eg2}
\end{align}
with $|\pm\rangle= (|e_1\rangle\pm|e_2\rangle)/\sqrt{2}$ such that $R$ is strongly monotonic. The initial condition is $\rho_0=|\psi_0\rangle\!\langle\psi_0|$ with $|\psi_0\rangle \propto ( |g\rangle+\alpha|e_1\rangle+\beta|e_2\rangle+\gamma|d\rangle)$ and parameters $\alpha,\beta,\gamma$.

\begin{figure}
\includegraphics[width=8.5cm]{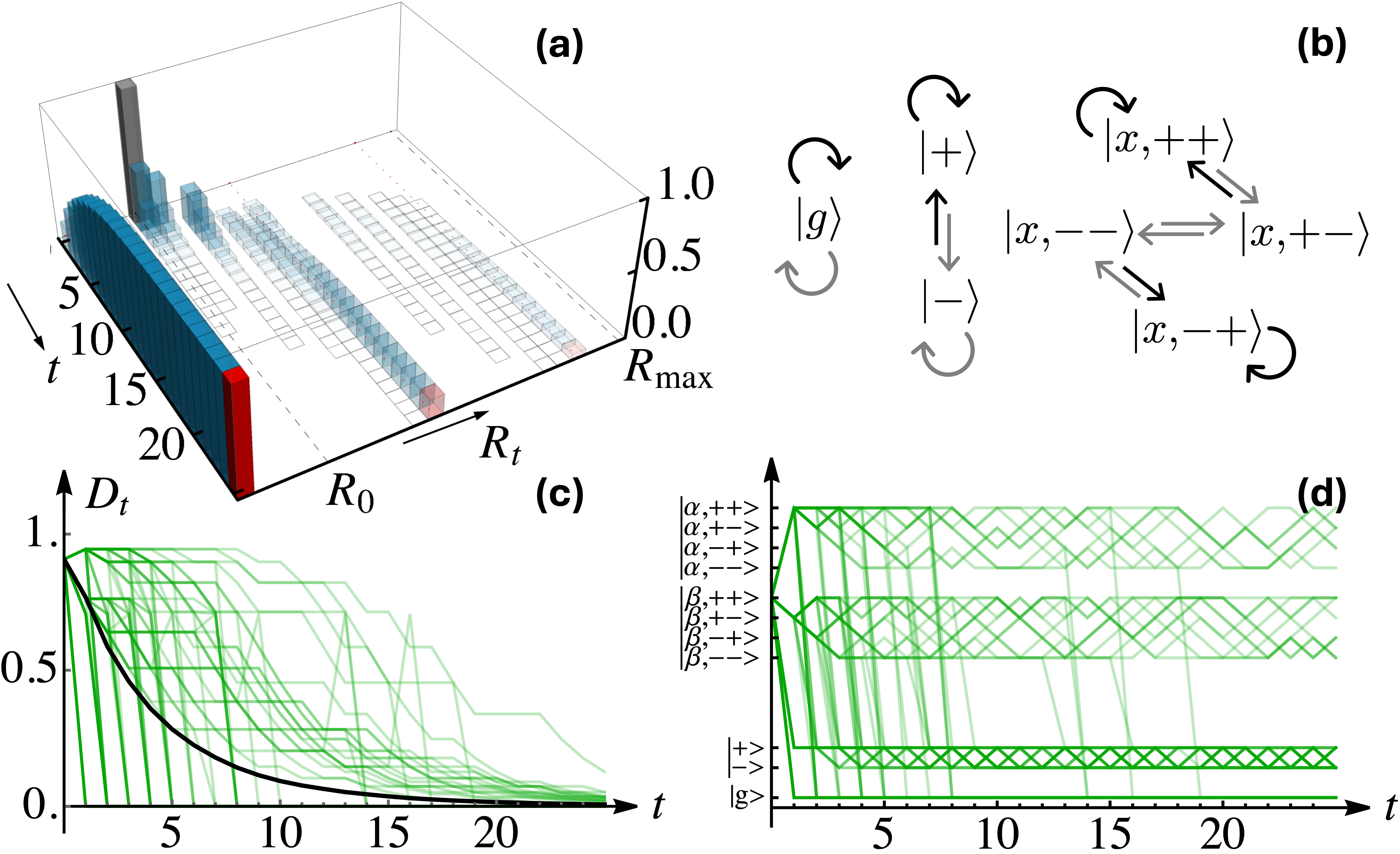}
\caption{\textbf{{Coherence freezing} in Example II.}  (a) Distribution of $R_t$ as a function of time, with the distribution of $R_\infty$ in red. (b) Schematic of the orbits to which $\varrho_t$ converges at large times.
The states comprising the four-state orbits are 
$|x,\pm,\sigma\rangle\!\langle x,\pm,\sigma|$
with 
$|x,\pm,\sigma\rangle\propto|g\rangle \pm \sqrt{2} x |\sigma\rangle$, where $x=\alpha$ or $\beta$ are determined by the initial state and $\sigma=\pm$.
(c) The trace distance $D_t$ of $\varrho_t$ to the nearest orbit for 100 trajectories (green) and the average of this quantity (black).
(d) The nearest orbit-state to $\varrho_t$.
}
\label{fig:orbits}
\end{figure}

Fig.~\ref{fig:orbits}(a) shows the probability distribution of $R_t$, as a function of time.  Here, $R_\infty$ is a random variable with three possible values (resource freezing).  Since $R$ is continuous, the conditional state approaches the relevant isosurface of $R$ at large times.  We summarise the behaviour here, with detailed analysis in~\cite{SM}.
The subchannel $\Lambda_4$ leaves the system in $|g\rangle\!\langle g|$, which is invariant under all $\Lambda_m$, so the state remains constant for all future times. It has no coherence, so this leads to $R_\infty=0$.  Subchannel $\Lambda_3$ leaves the system in $|e_1\rangle\!\langle e_1|$; after such an event, the system fluctuates between $|+\rangle\!\langle +|$ and $|-\rangle\!\langle -|$ and one has again $R_\infty=0$.  We refer to this fluctuating behaviour as a (stochastic) two-state \emph{orbit} (and we interpret the previous case as a one-state orbit).  Another possibility is that
neither $\Lambda_3$ or $\Lambda_4$ is ever applied (this does occur with positive probability); then the state converges to one of two possible four-state orbits, which have $R_\infty>0$.  Fig.~\ref{fig:orbits}(b) illustrates all the possible orbits (they depend in general on the parameters $\alpha,\beta$).
 
Hence, convergence of $\varrho_t$ to an isosurface of $R$ means that
$\varrho_t$ is always close {to} one of the orbit states at large times.  This is confirmed in Fig.~\ref{fig:orbits}(c), which shows that the {trace} distance to the nearest such state tends to zero along sampled {quantum} trajectories. 
Fig.~\ref{fig:orbits}(d) shows how the nearest state varies with time,  revealing the orbits.   Note (i)~for the four-state orbits, the state approaches the isosurface asymptotically, but never actually reaches it (unless $\gamma=0$, when a strong symmetry~\cite{Albert2014} arises from a weak symmetry~\cite{Buca2012}); (ii)~states that are close to a four-state orbit can change to the zero-resource orbit, but this has vanishing probability at large times.  These results illustrate how the long-time behavior of individual quantum trajectories is constrained by the martingale convergence theorem, with implications for the quantum state, as well as the distinction between vanishing and freezing.

\emph{Outlook}-- 
{By identifying strongly monotonic quantum resources with supermartingales, we proved {general results for the dynamics of resources and conditional states}.}
Eqs.~\eqref{eq:stopping} and~\eqref{eq:prob} show that for finite times, quantum resources cannot be increased by adaptive strategies using free channels.  
For free operations, the resource must either vanish or freeze asymptotically.  Under strong monotonicity, we showed that this holds for quantum trajectories, with probability 1 [Eq.~\eqref{eq:conv}].
This reveals
the results of~\cite{Munoz2019} for dynamics with strong symmetries as one example of a more general phenomenon, which is coherence vanishing~\cite{SM}. (The purification of~\cite{maassen2006} is not related to a purity QRT~\cite{Horodecki2003}, see~\cite{SM})}.
We also see that if a system is initialized in a state that forms part of  a stochastic orbit, the value of $R_t$ never changes, neither under the deterministic dynamics, nor in quantum trajectories.  This means that the resource is fully frozen, cf.~\cite{bromley2015,cianciaruso2015}.

Our results extend to continuous time dynamics, described by quantum master equations~\cite{Lindblad1976,Gorini1976}, for continuous resource monotones~\cite{SM}, because
both the optimal stopping theorem and martingale convergence theorem hold for continuous time (see, e.g.,~\cite{Roldan2023}).  Quantum trajectories are then obtained via continuous measurement~\cite{Belavkin1976,Gardiner_Zoller_2004}. 
Our results also shed light on
the general structure of asymptotic distributions, which remains an important open question.  Indeed, the discrete dynamics of Example II when applied at exponential waiting times leads to a  continuous-time dynamics that seems to contradict {Theorem 4} in~\cite{schmolke2025}, see~\cite{SM}.  We look forward to future work in this direction.

\vfill\eject

\newpage
\phantom{.}\newpage

\section*{Appendix}


\newcommand{\negskip}{\vspace{-8pt}}

\negskip

\subsection*{Quantum resource theory, supermartingales, and adaptive channels} 
\negskip

The specification of a QRT prescribes a set of free states and a set of free operations.  (This specification may be implicit in general but we assume here that both sets are well-defined.)  Any resource measure $R$ must vanish for free states and must be non-increasing under free channels.  If follows that if two states $\rho_1,\rho_2$ are interconvertible by free operations then $R(\rho_1)=R(\rho_2)$.
The definition of a convex resource measure is that for all {states} $\rho_1,\rho_2$ we have
\beq
	R({p_1} \rho_1 + {p_2} \rho_2 ) \leq {p_1} R(\rho_1) +{p_2} R(\rho_2), 
	\label{eq:convex-endmatt}
\eeq
where $p_1,p_2$ are probabilities.
Note that for  resource measures that are both convex and strongly monotonic, Eq.~\eqref{eq:strong-mono} can be generalised [using Eqs.~(\ref{eq:channel_subchannel},\ref{eq:mono})] to
 \beq
 R\left[ \Lambda(\rho) \right]  \leq  \sum_{m=1}^M p_m(\rho) \,R\!\left[ \frac{\Lambda_m(\rho)}{p_m(\rho)} \right] 
 \leq   R(\rho)  \quad \forall\;\rho  \; .
 \eeq

We show that: \emph{If $R$ is strongly monotonic and $R_0=R(\rho_0)<\infty$ then $R_t$ is a supermartingale.}
As discussed in the main text, strong monotonicity implies Eq.~\eqref{eq:supermart}.
This is the central property of a supermartingale, see Section 10.3 of Ref.~\cite{williamsBook,Liptser_2001,Roldan2023}.  To satisfy the definition, note that the relevant filtration is the natural one for the process $\varrho_t$ and $R$ is adapted to this filtration.  The resource $R$ is non-negative so $\mathbb{E}[|R_t|]=\mathbb{E}[R_t]$ and Eq.~\eqref{eq:supermart} shows that $\mathbb{E}[R_t]<\mathbb{E}[R_0]=R_0$ which we assumed to be finite.   Hence $\mathbb{E}(|R_t|)$ is bounded and all the conditions are satisfied for $R_t$ to be a supermartingale.

As a generalization of the setup from the main text, we note that $R_t$ is still a supermartingale if one applies {different strongly-monotonic channels} on each time step.
This also remains true if the channel that is applied on each step is chosen adaptively, based on {outcomes} collected in previous steps (see Sec 10.6 of~\cite{williamsBook}).  Specifically, {given} previous measurement outcomes $(m_1,m_2,\dots,m_{t})$,
one recovers a trajectory $(\rho_0;\varrho_1,\dots,\varrho_t)$ .  The resource $R_t$ is still a supermartingale if the channel that is applied in step $t+1$ depends on {that information (e.g., previous resource values)}, see~\cite{SM} for further discussion.

\negskip
\subsection*{Post-selection and {adaptive strategies}}
\negskip

We derive Eq.~\eqref{eq:stopping}  {for} a setup that includes the situation described in the main text, and also captures more general post-selection strategies.  
The initial state is $\rho_0$.  In step $t$, we choose adaptively a channel, previous measurement outcomes and resource values.  The stopping time $\tau$ is determined by a stopping criterion that also depends on this information.  We assume that the resource measure is strongly monotonic for all such channels, so $R_t$ is a supermartingale.
Then, the optional stopping theorem states that if the random time $\tau$ is bounded, we have (see Sec. 10.10 of \cite{williamsBook} or Theorem 10 of~\cite{Roldan2023}):
\beq
\mathbb{E}[ R_{\tau} ] \leq \mathbb{E}[ R_0 ] \, .
\label{eq:post-endmatt}
\eeq
The main text considers a non-adaptive channel with bounded $\tau\leq T$, so this result holds and Eqs.~(\ref{eq:stopping},\ref{eq:prob}) follow.
The probability of success is still $p^*=\mathbb{P}[R_\tau\geq R^*]$, so the fact that the optional stopping theorem holds also for adaptive channels means that improvements of the post-selection method of the main text cannot evade the bounds of Eqs.~(\ref{eq:stopping},\ref{eq:prob}).

Results similar to Eqs.~(\ref{eq:post-endmatt},\ref{eq:prob}) may also be proved under weaker conditions~\cite{williamsBook}.  
In particular, we show that: \emph{If $R$ is bounded then one may remove the cutoff $\Tmax$ and Eq.~\eqref{eq:prob} still holds.}  In this case, Eq.~\eqref{eq:post-endmatt} becomes (**)
\beq
\mathbb{E}[ R_{{\rm min}(\tau,\infty) ]
\leq \mathbb{E}[ R_0 ]}\, ,
\label{eq:post-inf}
\eeq
where $R_{{\rm min}(\tau,\infty)}$ is defined as $R_{{\rm min}(\tau,\infty)}=R_\tau$ if post-selection is successful ($\tau<\infty$) but $R_{{\rm min}(\tau,\infty)}=R_\infty = \lim_{t\to\infty} R_t $ if the stopping criterion is never achieved.  (The limit
exists almost surely because $R_t$ is a non-negative supermartingale, see below). 
Then Eq.~\eqref{eq:post-inf} is sufficient to establish the bound \eqref{eq:prob}, as required
(with success probability $p^*=\mathbb{P}[\tau<\infty]$).

 \begin{figure}
\includegraphics[width=8.5cm]{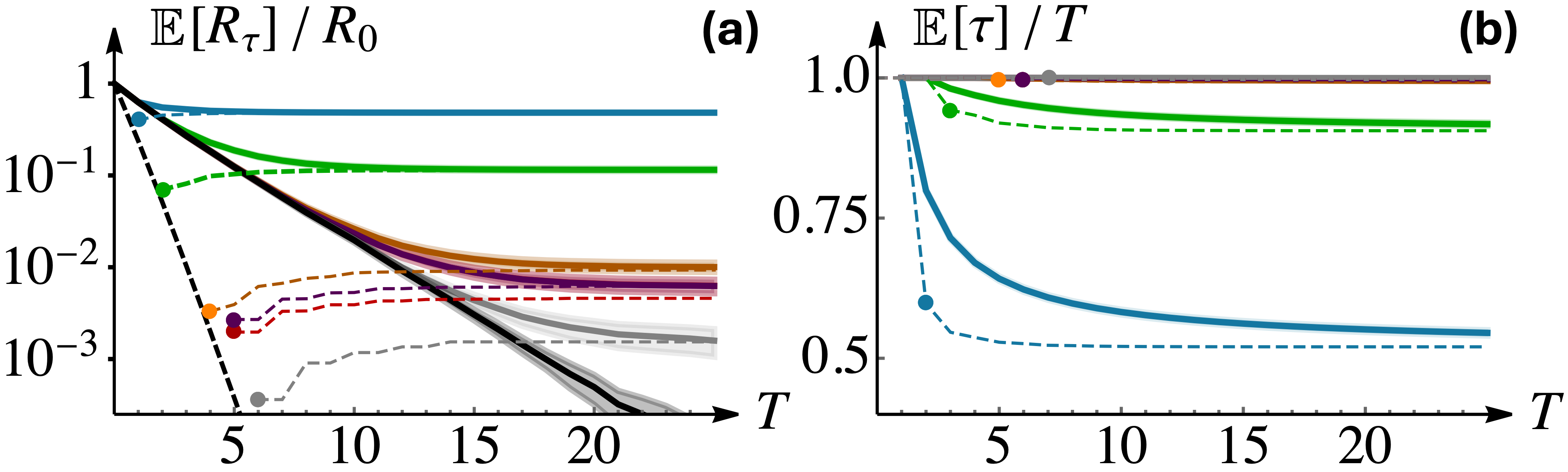}
\vspace{-0.2cm}
\caption{
{(a) Average post-selected resource $\mathbb{E}[R_\tau]$  in Example~I for selected thresholds $R^*$ (colour-coded as in Fig.~\ref{fig:post_selection}), as a function of the cutoff $T$.  The average without post-selection ($\mathbb{E}[R_T]$) is shown in black.  Dashed lines show {$ p^* R^*$}.   
(b)~The average stopping time measured relative to the cutoff: $\mathbb{E}[\tau]/T$; this is bounded from below by $\mathbb{P}[\tau=T]$, shown as dashed line.  Markers {in (a) and (b)} show the earliest times $T$ for which {$p^*>0$ and $\mathbb{P}[\tau=T]<1$, respectively}.
}}
\label{fig:cd}
\vspace{-8pt}
\end{figure}

\negskip
\subsection*{Cutoff {and threshold dependence in}  post-selection}
\negskip

The optional stopping theorem also holds more generally. Given two stopping criteria whose stopping times  satisfy $\tau_1\geq \tau_2$ with probability 1, we have (see Theorem 2.9 in~\cite{Liptser_2001} or Theorem 11~\cite{Roldan2023}):
\beq
\mathbb{E}[ R_{\tau_1} ] \leq \mathbb{E}[ R_{\tau_2} ].
\label{eq:post-endmatt-gen}
\eeq
Hence, if a post-selection strategy always stops later, its average resource is no larger. 
In particular, choosing $\tau_1=T$ and $\tau_2=\tau$ in Eq.~\eqref{eq:post-endmatt-gen}, one sees that
$\mathbb{E}[ R_{\tau}] \geq \mathbb{E}[ R_{T} ] $: the average resource under (general) post-selection with cutoff $T$ is no smaller than the average resource at {that} time (obtained without post-selection).

Furthermore, for (general) post-selection with two different cutoff times $T_1>T_2$, one has $\tau_1\geq \tau_2$.  Using this with Eq.~\eqref{eq:post-endmatt-gen}, one sees that $\mathbb{E}[ R_{\tau}] $ is non-increasing as a function of the cutoff time $T$, in contrast with the success probability $p^*$ (and $p^*R^*$).
The random time $\tau$ is also non-decreasing with $T$, so the average stopping time $\mathbb{E}[ \tau]$ is also non-decreasing.
Note also that $\mathbb{E}[\tau]\geq T\,\mathbb{P}[\tau=T]$, where we identify $\mathbb{P}[\tau=T]$ as the probability of not stopping before the cutoff {(that is, $\mathbb{P}[\tau=T]=1-\mathbb{P}[\tau<T]$). This probability $\mathbb{P}[\tau<T]$ is non-decreasing with $T$}, so $\mathbb{E}[\tau]/T$ is non-increasing.

Finally, consider post-selection with two different thresholds $R_1^* > R_2^*$.   Lower thresholds are crossed earlier so one has $\tau_1 \geq \tau_2$ and Eq.~\eqref{eq:post-endmatt-gen} applies.  Repeating the arguments from above, this means that {as $R^*$ grows}, the average post-selected resource {$\mathbb{E}[ R_{\tau}]$} generally decreases (and so does $p^*$, but not necessarily $p^*R^*$), while the average stopping time {$\mathbb{E}[ \tau]$} increases.
These results are illustrated in Fig.~\ref{fig:cd} with results for Example~I.

\negskip
\subsection*{Freezing, vanishing, and martingale convergence}
\negskip

We derive some properties of the large-time limit of $R_t$ that are quoted in the main text.
We show that: \emph{If $R$ is strongly monotonic then $R_t$ converges almost surely to a random variable $R_\infty$ at large times.  That is, the limit}
\beq
R_\infty = \lim_{t\to\infty} R_t  \quad \text{\emph{exists with probability $1$}}.
\label{eq:conv-end}
\eeq
Corollary 11.7 of~\cite{williamsBook} states that the above statement holds whenever $R_t$ is a non-negative supermartingale.  
We have showed above that strong monotonocity of $R$ is sufficient to ensure this, so the statement holds.  
(This is still true if the channels are chosen adaptively.)

Note $R_\infty$ is a random variable, it's value can be different for each trajectory.  
The definition of \emph{resource vanishing} is that $\mathbb{P}[R_\infty=0]=1$.  
The definition of \emph{resource freezing} is that $\mathbb{P}[R_\infty>0]>0$. 

We show that: \emph{If $R$ is strongly monotonic then the resource either vanishes or freezes}.  
Strong monotonicity implies Eq.~\eqref{eq:conv-end}; using also that 
$R$ is non-negative, one has 
\beq
\mathbb{P}[R_\infty=0] + \mathbb{P}[R_\infty>0]=1.
\eeq 
Since probabilities are non-negative, this means that either 
$\mathbb{P}[R_\infty=0]=1$ (vanishing) or $\mathbb{P}[R_\infty>0]>0$ (freezing), as required.

We show that: \emph{If $R$ is convex and bounded then resource vanishing implies that the resource of the average state vanishes at large times.} 
 Write $R^{\rm ave}_\infty = \lim_{t\to\infty} R(\rho_t)$ for the resource of the average state at larges times.
Using convexity [Eq.~\eqref{eq:convex-endmatt}] and the definition of the expectation implies 
\beq
R(\mathbb{E}[\varrho_t])  \leq \mathbb{E}\left[ R(\varrho_t) \right] .
 \label{eq:Rave-conv}
\eeq
Using Eq.~\eqref{eq:rho-ave} and 
taking the large time limit gives
\beq
R^{\rm ave}_\infty  \leq \lim_{t\to\infty} \mathbb{E}\left[ R_t\right] . 
 \label{eq:Rave-conv-large-t}
\eeq
Resource vanishing means that $\lim_{t\to\infty}R_t=0$ with probability $1$, which together with boundedness of $R$ means that $\lim_{t\to\infty}\mathbb{E}\left[ R_t \right] =0$.  Hence $R^{\rm ave}_\infty \leq 0$ and since $R$ is non-negative we have $R^{\rm ave}_\infty =0$, as required.

Similarly, we show that: \emph{If $R$ is convex and bounded and the average state has positive resource measure at large times, then the resource must freeze.} 
Convexity implies  Eq.~\eqref{eq:Rave-conv-large-t} so if the average state has positive resource measure ($R^{\rm ave}_\infty >0$) then also $\lim_{t\to\infty} \mathbb{E}\left[ R_t \right]>0$. 
Since $R$ is bounded then this implies that $R_\infty>0$ with non-zero probability, as required.

\negskip
\subsection*{Long-time behaviour of the conditional state}
\negskip

As noted in the main text, the fact that $R_t$ converges to a limit at long times has implications for the conditional state $\varrho_t$.  These are clearest if the resource measure $R$ is a continuous function, so we focus on this case.  
The definition of continuity requires a metric on the space of density matrices: the distance between $\rho,\rho'$  is denoted $d(\rho,\rho')$.  Write ${\cal S}_r$ for the set of density matrices $\rho$ that have $R(\rho)=r$. 
The distance of a state $\rho$ from $S_r$ is $D(\rho,r)=\inf_{\rho'\in {\cal S}_r} d(\rho,\rho')$.

We say that a trajectory is \emph{asymptotically supported on $S_r$} if $D(\varrho_t,r)\to 0$ at large times.  These sets are compact which means that asymptotic distribution of the state $\varrho_t$ is supported on \emph{cluster points}: that means that starting from a conditional state close to a cluster point, the trajectory will return arbitrarily close to the same point, an infinite number of times (see {also}~\cite{maassen2006}).

We show that: \emph{If $R$ is a continuous function of $\rho$ and $R_t$ is a supermartingale, then with probability 1, the trajectory is asymptotically supported on $S_{r_\infty}$, where $r_\infty$ is the value of $R_\infty$ on this trajectory.}
The martingale convergence theorem states that the random variable $R_t$ converges almost surely to $R_\infty$.  Let $r_\infty$ be the value of $R_\infty$ on the trajectory of interest: by continuity we have $D(\varrho_t,r_\infty)\to0$ as $t\to\infty$, as required.

We note that for some QRTs, it is possible to use as resource measure the function $R(\rho)={D}(\rho,0)$.  Then the set $S_r$ contains all states at a fixed distance from the set $S_0$  that contains all free states, from which trajectories with $R_\infty>r_\infty$ keep fixed distance $r_\infty$.

Finally, recall that a single resource may be quantified by two different measures $R,\tilde{R}$.  
 We show that: \emph{If two different measures $R,\tilde{R}$ are both continuous and faithful, then either they both freeze, or they both vanish.}
We give a proof by contradiction.  Suppose that $R$ vanishes and $\tilde{R}$ freezes.  Vanishing $R$ means that all trajectories have $R_\infty =0$; using this with results  from above, continuity of $R$ means that $\varrho_t$ is asymptotically supported on $S_0$.  Since $R$ is faithful, $S_0$ only contians free states, so $\tilde{R}(\varrho)=0$ for $\varrho\in S_0$.  Hence also $\lim_{t\to\infty}\tilde{R}_t=0$, and $\tilde{R}$ vanishes, contradicting the original assumption.  The analogous argument goes through if $R,\tilde{R}$ are swapped; hence either $R,\tilde{R}$ both freeze, or they both vanish.

\vfill\eject

\onecolumngrid
\newpage

\begin{center}
\textbf{Supplemental Material}
\end{center}


\renewcommand\thesection{S\arabic{section}}
\renewcommand\theequation{S\arabic{equation}}
\renewcommand\thefigure{S\arabic{figure}}
\setcounter{equation}{0}

\section{Quantum-Classical Measurement Maps (QCMMs) and Quantum Resource Theories}

Here we review the notion of QCMMs and how they relate to the concepts of strong-monotoniciy and convexity of resource measures in QRTs. 

\subsection{QCMMs}

The stochastic dynamics considered in the main text arises naturally in the context of QCMMs.  To explain this, we first introduce the notion of quantum-classical (QC) states.

We consider a joint density matrix for the system and a classical measurement register, of the form
\beq
\sigma^{\rm QC} = \sum_{m=1}^M p_m \sigma_m \otimes |m \rangle\!\langle m|,
\eeq
where $\sigma_m$'s are normalised states of the system, the $p_m$'s are  probabilities, and the $|m\rangle$ are elements of an orthonormal basis for the register, i.e., the classical basis.  This is referred to as a QC state,  as there are no coherences in the subsystem of the register. Such states arise on average from measurements with a discrete sets
of outcomes when $\sigma$ is converted to another
quantum state $\sigma_m$ with some probability $p_m$ and the corresponding outcome $m$ is recorded by ``appending'' a classical state $|m\rangle\!\langle m|$. 
Tracing the register out corresponds physically to discarding the register and is described a ``forgetting'' operation which is a linear map from QC states to quantum states of the system only:
\beq
 \sum_m p_m \sigma_m  \otimes |m \rangle\!\langle m| \mapsto  \sum_m p_m \sigma_m.
\eeq
(The range of summation is $1\dots M$, we leave this implicit in cases there is no ambiguity.)
By measuring the register in the classical basis before this operation, states $\sigma_m$'s can be recovered from the QC state.

We now connect to the stochastic dynamics of the main text. Given a decomposition into subchannels $S_\Lambda$ for a channel $\Lambda$, the corresponding QCMM is a linear operation from system density matrices to the QC states of the system and the register:
\beq
\varrho \mapsto \sum_m \Lambda_m(\varrho) \otimes |m \rangle\!\langle m| =\sum_m p_m(\varrho)\,\frac{\Lambda_m(\varrho)}{\Tr[\Lambda_m(\varrho)]} \otimes |m \rangle\!\langle m| ,
\label{eq:qmm-sub}
\eeq
where $p_m(\varrho)=\Tr[\Lambda_m(\varrho)]$.
This QCMM can be viewed as the average result of the action of a subchannel $\Lambda_m$  followed by appending the corresponding outcome (label) $m$. Applying the forgetting operation to the resulting QC state recovers the average state $\Lambda(\varrho)$, cf. Eq.~\eqref{eq:channel_subchannel}. If this operation is preceded by 
the measurement of the register in the classical basis, the (normalised) conditional state of the system is recovered, cf. Eq.~\eqref{eq:stochastic-state-evol}.  

We define QCMMs that store a history of measurement outcomes by acting on joint QC states.  Write 
\beq
|\bm{m}_t\rangle=|m_t\rangle\otimes|m_{t-1}\rangle\otimes\cdots \,|m_1\rangle
\eeq
for a state vector that represents the sequence of measurement outcomes up to time $t$; similarly $\bm{m}_t=(m_1,\dots,m_t)$.
Then a QCMM that stores the history is 
\beq
\varrho_t \otimes |\bm{m}_t\rangle\!\langle \bm{m}_t|
\mapsto \sum_{m_{t+1}=1}^M \Lambda_{m_{t+1}} (\varrho_t)  \otimes  |m_{t+1}\rangle\!\langle m_{t+1}|\otimes |\bm{m}_t \rangle\!\langle \bm{m}_t|.
\label{equ:qmm-record}
\eeq
where we identify the RHS as $\sum_{m_{t+1}} \Lambda_{m_{t+1}}(\varrho_t)  \otimes  |\bm{m}_{t+1} \rangle\!\langle \bm{m}_{t+1}|$,

\subsection{QRTs with QCMMs}

We now show that in QRTs where a QCMM is allowed, strong monotonicity and convexity follow for the corresponding operation on the system under modest additional assumptions.  These arguments follow closely the review of~\cite{chitambar2019rmp}. 

Consider a QRT where a QCMM  as in Eq.~\eqref{eq:qmm-sub} is free. 
We also require the resource measure $R$ to be well-defined both on the space of the system and on the joint space of the system and the register.  
Since the QCMM of Eq.~\eqref{eq:qmm-sub} is free, it cannot increase the resource, so 
\begin{equation}
 R\left[\sum_m \Lambda_m(\varrho) \otimes |m \rangle\!\langle m| \right] \leq  R(\varrho).\label{eq:sm_r1}
\end{equation}

In addition to Eq.~\eqref{eq:qmm-sub} being free, we make two assumptions on the QRT.  The first is that measuring the register in the classical basis is free. Then, the average of the resource over the conditional states is not greater than that of the QC state:
\begin{equation}
\sum_m p_{m}(\varrho) \,R\left\{ \frac{\Lambda_m(\varrho)}{\Tr[\Lambda_m(\varrho)]} \otimes |m \rangle\!\langle m|\right\}   \leq   R\left[\sum_m \Lambda_m(\varrho) \otimes |m \rangle\!\langle m| \right].\label{eq:sm_r2}
\end{equation}
The second assumption is
that forgetting outcomes is free. {We can then} discard the register and the resource does not increase for each conditional state:
\begin{equation}
\sum_m p_{m}(\varrho) \,R\left\{ \frac{\Lambda_m(\varrho)}{\Tr[\Lambda_m(\varrho)]} \right\}  \leq \sum_m p_{m}(\varrho) \,R\left\{ \frac{\Lambda_m(\varrho)}{\Tr[\Lambda_m(\varrho)]} \otimes |m \rangle\!\langle m|\right\}. \label{eq:sm_r3}
\end{equation}
Combining Eqs.\eqref{eq:sm_r1}-\eqref{eq:sm_r3}, we arrive at 
\begin{equation}
\sum_m p_{m}(\varrho) \,R\left\{ \frac{\Lambda_m(\varrho)}{\Tr[\Lambda_m(\varrho)]} \right\} \leq R(\varrho), \label{eq:sm_r4}
\end{equation}
so indeed $R$ is strongly monotonic under these assumptions [cf. Eq.~\eqref{eq:strong-mono}].
Also, by discarding the register for the QC state~\eqref{eq:qmm-sub}, we recover from Eq.~\eqref{eq:sm_r1} the monotonicity of $R$ {[cf. Eq.~\eqref{eq:mono}]},
\begin{equation}
\,R\left[\Lambda(\varrho) \right] \leq R(\varrho). \label{eq:sm_r5}
\end{equation} 

We make two remarks on the assumptions required to derive Eq.~\eqref{eq:sm_r4}. 
First, it may be that the resource measure $R$ is a convex-linear function: this is the condition that  Eq.~\eqref{eq:sm_r2} holds for all (sets of) subchannels~\cite{chitambar2019rmp}.  In this case we can drop the assumption that measuring the register is free.
Second, we assumed that forgetting is free; if we also assume that appending a classical register is free (in an arbitrary classical state), then it can be shown that the resource measure is unchanged by forgetting~\cite{chitambar2019rmp}, so that Eq.~\eqref{eq:sm_r3} becomes an equality.

Moreover, if $R$ is indeed convex-linear (on the joint QC space), then its action on the system state alone is constrained as
\begin{equation}
\,R\left[\Lambda(\varrho) \right] \leq \sum_m p_{m}(\varrho) \,R\left\{ \frac{\Lambda_m(\varrho)}{\Tr[\Lambda_m(\varrho)]} \right\} , \label{eq:sm_r6}
\end{equation}
which holds for any decomposition of an arbitrary channel $\Lambda$.
That is, $R$ is convex for ensembles generated by measuring and forgetting the classical part of the QC states.

\subsection{Adaptive QCMMs and relation to post-selection}

As discussed in the main text, it is natural in the context of martingale theory to consider adaptive channels where the operation at step $t+1$ is chosen according to previous measurements.

An \emph{adaptive QCMM} is a measurement where the subchannels at step $t+1$ depend on the previous measurement outcomes.  Writing $M_{t+1}$ for the number of subchannels in this step, the adaptive QCMM may be written as
\beq
\varrho_t \otimes |\bm{m}_t\rangle\!\langle \bm{m}_t|
\mapsto
\sum_{m_{t+1}=1}^{M_{t+1}} \Lambda_{m_{t+1}}^{(\bm{m}_t)} (\varrho_t)  \otimes  |\bm{m}_{t+1} \rangle\!\langle \bm{m}_{t+1}|,
\eeq
subject to the constraint that each {$\Lambda_{m_{t+1}}^{(\bm{m}_t)}$ is CP and $\Lambda^{(\bm{m}_t)}=\sum_{m_{t+1}} \Lambda_{m}^{(\bm{m}_t)}$} is CPTP.
Note that since $\rho_0$ is fixed then the sequence $\bm{m}_t$ fixes the conditional state at time $t$ as
\beq
\varrho_t \,{\propto} \,\Lambda^{(\bm{m}_t)}_{m_t} \circ \cdots \circ  \Lambda^{(\bm{m}_2)}_{m_2} \circ \Lambda^{(\bm{m}_1)}_{m_1}(\rho_0),
\eeq 
where the $\circ$ indicates composition of operators.  That means that $R_t$ can be expressed as a function of $\bm{m}_t$. 

Such adaptive channels are natural in the when considering post-selection together with martingale theory, since they provide an alternative framework in which to implement stopping times.  For example, the post-selection method described in the main text can be alternatively implemented by an adaptive channel as:
\beq
 (\Lambda_{m}^{(\bm{m}_t)},M_{t+1}) = \begin{cases} (\Lambda_m,M) & \quad \text{if $R_t<R^*$}
  \\
  (\Lambda_{\bf{1}},1) & \quad \text{otherwise}. \end{cases}
\eeq
where $\Lambda_{\bf{1}}$ is the identity channel, that is $\Lambda_{\bf{1}}(\varrho)=\varrho$ for all $\varrho$.  (This channel is free in all QRTs and it leaves $R_t$ unchanged.)  Physically, this adaptive channel means that one applies the QCMM of Eq.~\eqref{eq:qmm-sub} until the resource threshold is achieved, after which one applies the identity so that the conditional state no longer changes.

Then, the post-selection method of the main text (with cutoff $T$) simply corresponds  to $T$ iterations of the adaptive channel: we have rephrased the idea of a stopping criterion in the language of QCMMs.  The post-selection is successful if $R_T\geq R^*$ in which case the final state $\varrho_T$ has the desired resource; otherwise $R_T<R^*$ and  the attempt failed.  
The same arguments are also applicable to the generalized post-selection method described in the Appendix to the main text.

\section{Further discussion of examples}

\subsection{Example I}

\emph{Channel and subchannels}.  We consider a system of two qubits as in Fig.~\ref{fig:illustrate}(a).  We introduce three subchannels $\Lambda_m(\rho)=K_m \rho K_m^\dag$ with Kraus operators on the two-qubit space:
\begin{subequations}
\begin{align}
K_{1}&=\kappa(\theta)\otimes \kappa(\theta),\\
K_{2}&=\kappa(\theta)\otimes \kappa(\theta+\pi/2),\\
K_{3}&=\kappa(\theta+\pi/2)\otimes \upsilon ,
\end{align}
\end{subequations}
where $\theta$ is a parameter and 
\beq
\kappa(\theta)=\cos(\theta)|\!\!\uparrow\rangle\!\langle\downarrow\!\!| + \sin(\theta)|\!\!\downarrow\rangle\!\langle\uparrow\!\!|,\qquad \upsilon=|\!\!\uparrow\rangle\!\langle\downarrow\!\!| + |\!\!\downarrow\rangle\!\langle\uparrow\!\!|,
\eeq
{are one-qubit operators; also
 $\upsilon={\kappa(\pi/4)\sqrt{2}}$}.  Note that $\kappa$ is non-unitary in general but $\upsilon$ is unitary.
We consider the channel $\Lambda(\rho)=\sum_{m=1}^3 \Lambda_m(\rho)$ as in Eq.~\eqref{eq:channel_subchannel}.
Since each subchannel is constructed from a single Kraus operator, the initial pure state $\rho_0=|\psi_0\rangle\!\langle \psi_0|$ remains pure for all times along quantum trajectories.
As noted in the main text, we take $|\psi_0\rangle=(|00\rangle+|01\rangle+|11\rangle)/\sqrt{3}$.

{Special cases occur when $\theta=n\pi/4$ for integer $n$.  For $n$ odd, all Kraus operators are proportional to unitaries and the entanglement remains always constant along quantum trajectories.  For $n$ even, the conditional state is a product state and the entanglement is zero for all times.}
{Outside of these special cases, the entanglement has non-trivial stochastic dynamics.  The results of the main text have $\theta=\arccos(1/\sqrt{5})$.}

\smallskip
\emph{Interpretation as an LOCC operation}.  As discussed in the main text, the channel can seen explicitly as an LOCC operation, by breaking it down into three steps.   The procedure is naturally expressed using an adaptive QCMM, {which highlights the adaptive structure of that set of free operations}.
We define local operators $\kappa_A(\theta)=\kappa(\theta)\otimes \mathbb{1}$ and $\kappa_B(\theta)=\mathbb{1}\otimes \kappa(\theta)$ that act on qubits $A$ and $B$ individually.
The first step is a LO on qubit $A$ that includes a generalised measurement, which we express as
\beq
\varrho \mapsto \kappa_A(\theta) \varrho \kappa_A(\theta)^\dag \otimes | 1_A\rangle\!\langle 1_A|
+
\kappa_A(\theta+\pi/2) \varrho \kappa_A(\theta+\pi/2)^\dag \otimes | 2_A\rangle\!\langle 2_A|,
\label{eq:ex1-lo1}
\eeq
where the measurement outcomes are labelled as $1_A$ and $2_A$.  The measurement outcome is communicated classically to qubit $B$.  Then one applies a LO on $B$, dependent on the result of the previous measurement.  This can be expressed an adaptive QCMM similar to Eq.~\eqref{equ:qmm-record}: If the first outcome is $1_A$ then
\begin{equation}
\varrho\otimes | 1_A\rangle\!\langle 1_A|  
\;\mapsto\;
\kappa_B(\theta) \varrho \kappa_B(\theta)^\dag \otimes | 1_A1_B\rangle\!\langle 1_A1_B|
+
\kappa_B(\theta+\pi/2) \varrho \kappa_B(\theta+\pi/2)^\dag \otimes | 1_A2_B\rangle\!\langle 1_A2_B|,
\end{equation}
where one sees that the result of the second measurement is appended to the register.
Alternatively, if the first outcome is $2_A$ then the second step is
\begin{equation}
\varrho\otimes | 2_A\rangle\!\langle 2_A| 
\;\mapsto\;
\upsilon_B \varrho \upsilon_B^\dag \otimes |2_A {1_B}\rangle\!\langle 2_A {1_B}|,
\label{eq:ex1-lo2}
\end{equation}
with $\upsilon_B=\mathbb{1}\otimes\upsilon$.
In this case the operation is unitary, so {a single} measurement outcome is appended.
Noting that $K_1=\kappa_B(\theta)\kappa_A(\theta)$ and $K_2=\kappa_B(\theta+\pi/2)\kappa_A(\theta)$ and $K_3=\upsilon_B\kappa_A(\theta+\pi/2)$
the composition of the QCMMs for the two steps can be expressed as 
\beq
\varrho \;\mapsto\; \sum_{m=1}^3 K_m \varrho K_m^\dag \otimes |m\rangle\!\langle m|,
\eeq
where the three possible outcomes  of the adaptive QCMM $(1_A1_B,1_A2_B,2_A {1_B})$, have been relabelled as $m=(1,2,3)$ respectively.  Comparing with Eq.~\eqref{eq:qmm-sub}, this final QCMM realises the channel of Example I, including its decomposition into subchannels.

\smallskip
\emph{Resource Measure}. We choose the entanglement of formation~\cite{Bennett1996b}, which is faithful, convex, and strongly monotonic with respect to LOCC channels.  For pure states the entanglement of formation coincides with the {so-called} entropy of entanglement~\cite{Horodecki2003}. That is, for a system consisting of two subsystems $A$ and $B$, the entanglement is 
\begin{equation}
R(|\psi\rangle\!\langle \psi|)=S_{\mathrm{vN}}[ \Tr_B(|\psi\rangle\!\langle \psi|)],
\end{equation}
where $\Tr_B$ denotes the partial trace over the subsystem $B$ and $S^{(\mathrm{vN})}$ is the von Neumann entropy. 
For a pure state of 2 qubits as in Example I, the entanglement can be expressed using its coefficients in the basis $|\!\uparrow\uparrow\rangle$, $|\!\uparrow\downarrow\rangle$, $|\!\downarrow\uparrow\rangle$, $|\!\downarrow\downarrow\rangle$, as
\begin{equation}
R(|\psi\rangle\!\langle \psi|)=S_{\rm mix}
\!\left[\frac{1+\sqrt{(|\psi_{\!\uparrow\uparrow}|^2+|\psi_{\!\uparrow\downarrow}|^2-|\psi_{\!\downarrow\uparrow}|^2-|\psi_{\!\downarrow\downarrow}|^2)^2+4|\psi_{\!\uparrow\uparrow}\psi^*_{\!\downarrow\uparrow}+\psi_{\!\uparrow\downarrow}\psi^*_{\!\downarrow\downarrow}|^2 }}{2}\right], 
\end{equation}
where $S_{\rm mix}(x)=-x \log x-(1-x) \log (1-x) $. For the initial state $\rho_0$ 
the entanglement is $R(\rho_0)=S_{\rm mix}[(3+\sqrt{5})/6]=0.381...$.

\smallskip
\emph{Average Dynamics}. For the generic case $\theta\neq n\pi/4$
the average dynamics features a periodic orbit consisting of 2 states:
\begin{subequations}
\begin{align}
\rho_{\infty;+}&= p_1 \,|\!\!\uparrow\rangle\!\langle \uparrow\!\!| \otimes|\!\!\uparrow\rangle\!\langle \uparrow\!\!| 
+p_2\,|\!\!\uparrow\rangle\!\langle \uparrow\!\!| \otimes|\!\!\downarrow\rangle\!\langle \downarrow\!\!|
+p_3\,|\!\!\downarrow\rangle\!\langle \downarrow\!\!| \otimes|\!\!\uparrow\rangle\!\langle \uparrow\!\! |
+p_4\,|\!\!\downarrow\rangle\!\langle \downarrow\!\!| \otimes|\!\!\downarrow\rangle\!\langle\downarrow \!\!|,\\
\rho_{\infty;-}&= p_1 \,|\!\!\downarrow\rangle\!\langle \downarrow\!\!| \otimes|\!\!\downarrow\rangle\!\langle\downarrow \!\!|
+p_2\,|\!\!\downarrow\rangle\!\langle \downarrow\!\!| \otimes|\!\!\uparrow\rangle\!\langle \uparrow\!\! |
+p_3\,|\!\!\uparrow\rangle\!\langle \uparrow\!\!| \otimes|\!\!\downarrow\rangle\!\langle \downarrow\!\!|
+p_4\,|\!\!\uparrow\rangle\!\langle \uparrow\!\!| 
\otimes|\!\!\uparrow\rangle\!\langle \uparrow\!\!|,
\end{align}
\end{subequations}
where the probabilities $p_1$,$p_2$, $p_3$, $p_4$ depend on the initial condition. 
That is, 
$\Lambda( \rho_{\infty;\pm} ) = \rho_{\infty;\mp}$
(there are {four} eigenvalues of $\Lambda$ with unit modulus: {$1,-1,1,-1$ corresponding to $|\!\!\uparrow\rangle\!\langle \uparrow\!\!|\otimes |\!\!\uparrow\rangle\!\langle \uparrow\!\!|+|\!\!\downarrow\rangle\!\langle \downarrow\!\!|\otimes|\!\!\downarrow\rangle\!\langle \downarrow\!\!|$, $|\!\!\uparrow\rangle\!\langle \uparrow\!\!|\otimes |\!\!\uparrow\rangle\!\langle \uparrow\!\!|-|\!\!\downarrow\rangle\!\langle \downarrow\!\!|\otimes|\!\!\downarrow\rangle\!\langle \downarrow\!\!|$, $|\!\!\uparrow\rangle\!\langle \uparrow\!\!|\otimes |\!\!\downarrow\rangle\!\langle \downarrow\!\!|+|\!\!\downarrow\rangle\!\langle \downarrow\!\!|\otimes|\!\!\uparrow\rangle\!\langle \uparrow\!\!|$, and $|\!\!\uparrow\rangle\!\langle \uparrow\!\!|\otimes |\!\!\downarrow\rangle\!\langle \downarrow\!\!|-|\!\!\downarrow\rangle\!\langle \downarrow\!\!|\otimes|\!\!\uparrow\rangle\!\langle \uparrow\!\!|$, respectively).

For mixed states of $2$ qubits, the entropy of formation $R$ can be calculated using the so-called concurrence~\cite{Wootters1998}. For either of the two states $\rho_{\infty;+}$  and $\rho_{\infty;-}$, the entanglement vanishes,
\begin{equation}
R(\rho_{\infty;\pm})=0.
\end{equation}
This can be also seen directly as these states are separable as probabilistic mixtures of product states.
}

\subsection{Example II}

\emph{Channel and subchannels}.  As described in the main text we consider a system with four (orthonormal) basis states $|g\rangle$, $|e_1\rangle$, $|e_2\rangle$, $|d\rangle$.   Recall that $|\pm\rangle= (|e_1\rangle\pm|e_2\rangle)/\sqrt{2}$.  There are four subchannels, each of which involves a single Kraus operator, see Eq.~\eqref{eq:kraus-eg2}. 
For a pure initial state (as considered in main text), this means that the conditional state remains pure for all times.

\smallskip
\emph{Resource measure}. We choose the relative entropy of coherence~\cite{vaccaro2008,baumgratz2014}, {which is faithful, convex, and strongly monotonic with respect to incoherent operations (e.g., subchannels with weak symmetry~\cite{Buca2012} {as considered here}). For a state $\rho$ it is defined as:}
\begin{equation}
R(\rho)=-S_{\mathrm{vN}}(\rho)+S_{\mathrm{vN}}[\Delta(\rho)],
\end{equation}
where 
$\Delta(\rho)$ -- which we refer to as the dephased state -- is obtained from $\rho$ by removing coherences between the ground state $|g\rangle$ and the rest {($|e_1\rangle$, $|e_2\rangle$, $|d\rangle$)}. That is, $\Delta(\rho)=\mathbb{1}_{Q_g}\rho\mathbb{1}_{Q_g}+(\mathbb{1}-\mathbb{1}_{Q_g})\rho(\mathbb{1}-\mathbb{1}_{Q_g})$,
where $\mathbb{1}_{Q_g}=|g\rangle\!\langle g|$ and $\mathbb{1}$ is the identity matrix. {Note that the coherence we are interested here (between the ground state and the rest of the basis states) is example of degenerate coherence~\cite{streltsov2017}}.

\smallskip
{\emph{Orbits}.  We describe the orbits to which the conditional state converges at long times.  We consider the general case where $\rho_0$ may be mixed (the main text specialises to pure $\rho_0$).
We write the matrix elements of the initial state as $\rho_{ij} = \langle i | \rho_0 |j\rangle$, where $|i\rangle,|j\rangle$ are basis states.

As noted in the main text, the state $|d\rangle$ {spans the} decay subspace, so initial states with $\rho_{dd}=0$ never acquire any support on this state.  For such initial states, subchannels $\Lambda_3,\Lambda_4$ never occur and the asymptotic fate of the system is controlled by the first time step.  If subchannel $\Lambda_1$ occurs in this first step then the system arrives immediately in a four-state orbit with states 
\beq
\omega_{1;\pm,\sigma}= \frac{\rho_{gg}\,|g\rangle\!\langle g| \pm \sqrt{2}\left(\rho_{g e_1}\,|g\rangle\!\langle \sigma|+\rho_{e_1 g}\,|\sigma\rangle\!\langle g|\right)+2\rho_{e_1 e_1}\,|\sigma\rangle\!\langle \sigma| }{\rho_{gg}+2\rho_{e_1 e_1}},
\label{eq:sm:ex2_omega_1}
\eeq
where $\sigma=+,-$.  On the other hand, if the subchannel $\Lambda_2$ occurs in the first step then the system arrives in a different {four-state} orbit whose states are
\beq
\omega_{2;\pm,\sigma}=\frac{\rho_{gg}\,|g\rangle\!\langle g| \pm \sqrt{2}\left(\rho_{g e_2}\,|g\rangle\!\langle \sigma|+\rho_{e_2 g}\,|\sigma\rangle\!\langle g|\right) +2\rho_{e_2 e_2}\,| 
\sigma\rangle\!\langle \sigma|}{\rho_{gg}+2\rho_{e_2 e_2}}.
\label{eq:sm:ex2_omega_2}
\eeq

If on the other hand $\rho_{dd}>0$, then {either} $\Lambda_3$ {or} $\Lambda_4$ may occur.  If $\Lambda_3$ ever happens, then it leaves the system in $|e_1\rangle\!\langle e_1|$; for all future times, the support on the decay subspace is zero and only subchannels $\Lambda_1,\Lambda_2$ can occur.  Whichever one occurs, the system arrives in a two-state orbit whose states are
\beq
\omega_{3;+} = |+\rangle\!\langle +|, \qquad \omega_{3;-} = |-\rangle\!\langle -| .
\label{eq:sm:ex2_omega_3}
\eeq
Alternatively if $\Lambda_4$ ever happens, then it leaves the system in $|g\rangle\!\langle g|$; the support on the decay subspace is again zero and only subchannels $\Lambda_1,\Lambda_2$ can occur after this.  These leave the state invariant so the system remains forever in the single-state orbit
\beq
\omega_{4}= |g\rangle\!\langle g|.
\label{eq:sm:ex2_omega_4}
\eeq
It is important that for $\rho_{dd}>0$ there is still a positive probability that neither $\Lambda_3$ nor $\Lambda_4$ ever occurs (see below).  In this case the system asymptotically approaches one of the four-state orbits.  (It approaches $\omega_1$ if the first subchannel was $\Lambda_1$ and $\omega_2$ if the first subchannel was $\Lambda_2$.)
The {stochastic dynamics} on $4$ possible orbits {above} are {analogous to those for pure initial states in} Fig.~\ref{fig:orbits} of the main text.

The probabilities of asymptotically approaching each orbit can be worked out as follows.  We write $p_m$ for the probability to approach the $m$th orbit $\omega_m$.  
Recall that the first orbit can only be reached if $\Lambda_1$ happens in the first step.  After this we require that neither $\Lambda_3$ nor $\Lambda_4$ ever happens.  {The case of the second orbit is analogous but with $\Lambda_2$ happening first.}
From the construction of the stochastic dynamics [Eq.~\eqref{eq:stochastic-state-evol}], {the corresponding probabilities are}
\begin{align}
p_1 & = \lim_{t\to\infty} \Tr[(\Lambda-\Lambda_3-\Lambda_4)^t \circ \Lambda_1 (\rho_0)] , \qquad p_2  = \lim_{t\to\infty} \Tr[(\Lambda-\Lambda_3-\Lambda_4)^t \circ \Lambda_2 (\rho_0)],
\end{align}
where $\circ$ indicates composition of the channels. These expressions can be evaluated to give
\beq
p_1=\frac{\rho_{gg}}{2}+\rho_{e_1 e_1}, \qquad
p_2=\frac{\rho_{gg}}{2}+\rho_{e_2 e_2} .
\eeq
We recall that $p_3$ is the probability that $\Lambda_3$ does occur, and similarly $p_4$ is the probability that $\Lambda_4$ does occur.   Noting that the probability that $\Lambda_3$ never occurs is $\lim_{t\to\infty} \Tr[(\Lambda-\Lambda_3)^t (\rho_0)]$ (and {analogously} for $\Lambda_4$), one has
\begin{align}
p_3 = 1 - \lim_{t\to\infty} \Tr[(\Lambda-\Lambda_3)^t (\rho_0)]
, \qquad
p_4 = 1 - \lim_{t\to\infty} \Tr[(\Lambda-\Lambda_4)^t (\rho_0)],
\end{align}
from which we obtain
\beq
p_3 = \frac{\rho_{dd}}{2} , \qquad  p_4 = \frac{\rho_{dd}}{2} .
\eeq

These probabilities can be understood as follows. The probability of neither $K_3$ nor $K_4$ ever happening is the probability of not being found in the decay subspace, $1-\rho_{dd}$. Then, $p_1$ and $p_2$ are equal to the (first-step) rates  for $K_1$ and $K_2$ over the non-decay subspace only. The probability for either of $K_3$ and $K_4$ to happen  is equally likely, so that $p_3=p_4$ [as their (any-step) rates over that subspace are equal], while they must sum $\rho_{dd}$.

Finally, we note that all the orbits have constant coherence, as required by Eq.~\eqref{eq:conv}. Indeed,  
\begin{subequations}
\begin{align}
R(\omega_{1;\pm,\sigma})&=-S_\mathrm{mix}\!\left[\frac12 + \frac{\sqrt{(\rho_{gg}-2\rho_{e_1 e_1})^2+4\rho_{ge_1}\rho_{e_1 g}} }{2(\rho_{gg}+2\rho_{e_1 e_1})}\right]+S_\mathrm{mix}\!\left(\frac{\rho_{gg}}{\rho_{gg}+2\rho_{e_1 e_1}}\right), \\
R(\omega_{2;\pm,\sigma})&=-S_\mathrm{mix}\!\left[\frac12 +\frac{\sqrt{(\rho_{gg}-2\rho_{e_2 e_2})^2+4\rho_{ge_2}\rho_{e_2 g}} }{2(\rho_{gg}+2\rho_{e_1 e_1})}\right]+S_\mathrm{mix}\!\left(\frac{\rho_{gg}}{\rho_{gg}+2\rho_{e_2 e_2}}\right),\\ 
R(\omega_{3;\pm})&=0, \\
R(\omega_{4})&=0. 
\end{align}
\end{subequations}
Recalling that $S_\mathrm{mix}(x)=S_\mathrm{mix}(1-x)$,
we  conclude that coherence freezing occurs if and only if $\rho_{ge_1}\rho_{e_1 g}\neq 0$ or $\rho_{ge_2}\rho_{e_2 g}\neq 0$, 
i.e., some coherences are initially present between the ground state $|g\rangle$ and the excited states $|e_1\rangle$, $|e_2\rangle$. Initial coherences to the decay state $|d\rangle$ do not play any role. 
\smallskip

\emph{Average dynamics}. The channel $\Lambda$ leaves two orthogonal states invariant: $|g\rangle\!\langle g|$ and $(|e_1\rangle\!\langle e_1|+|e_2\rangle\!\langle e_2|)/2 $, so that the average state with a general initial condition $\rho_0$ is asymptotically given by
\begin{equation}
\rho_\infty=p_g\,|g\rangle\!\langle g|+(1-p_g)\frac{|e_1\rangle\!\langle e_1|+|e_2\rangle\!\langle e_2|}{2},
\label{eq:sm:ex2_rho_ss}
\end{equation}
where $p_g=\Tr(L\rho_0)$ with $L=|g\rangle\!\langle g|+|d\rangle\!\langle d|/2$. 
The presence of a conservation law with respect to the observable $L$ is related to the strong symmetry~\cite{Albert2014} arising when restricting only to the non-decay subspace. The non-decay subspace is given by the maximal support of a general asymptotic state. That support is $Q_g\oplus Q_e$, where $Q_g$ and $Q_e$ are the supports of the two orthogonal invariant states, which are the subspaces spanned by $|g\rangle$ and $|e_1\rangle$, $|e_2\rangle$, respectively. Those are also the symmetry subspaces, in which Kraus operators are block-diagonal when restricted to the non-decay subspace.  

Note that the asymptotic state always has zero resource, that is,
\begin{equation}
R(\rho_\infty)=0,
\end{equation}
since there it features no coherence between subspaces $Q_g$ and $Q_e$. 
This means that the coherence vanishes in the average dynamics, even through it freezes in (stochastic) quantum trajectories.
This can be attributed to the asymptotic stochastic dynamics over orbits introducing random phases to such coherences which on average leads to their dephasing, cf. Fig.~\ref{fig:orbits}.     

\section{Relation to previous work}

We now explain how the resource convergence in Eq.~\eqref{eq:conv} is related to previously reported phenomena that can arise in quantum trajectories.  

\subsection{Dissipative freezing}

The phenomenon of \emph{dissipative freezing}~\cite{Munoz2019} has been observed for continuous dynamics with a strong symmetry, but can be also considered for discrete dynamics. In that case, each Kraus operator shares a unitary symmetry, which also implies that corresponding charge is conserved~\cite{Albert2014}. Thus, when a quantum trajectory is initialised within a single symmetry subspace, it remains there at all times. However, if it starts in a superposition over different symmetry subspaces, it may localise onto a single random subspace at long times: this is referred to as dissipative freezing. 

The relation to our work is as follows: Quantum channels with a strong symmetry are a subset of free operations called incoherent free operations in the resource theory of coherence.  The relative entropy of coherence is strongly monotonic for incoherent free operations.  Hence, our results show that either coherence vanishes asymptotically, or it freezes.  That is, for initial states with non-zero coherence (which are all asymmetric states), their coherence asymptotically either vanishes or approaches a non-zero constant limit. In particular, for asymmetric pure states (which remain pure under the stochastic dynamics), coherence vanishing means that  they are asymptotically supported on a single symmetry subspace (i.e., they become symmetric).  Coherence freezing means that they approach asymptotically some orbits that consist of superpositions over different symmetry subspaces.  These two cases (vanishing/freezing) are the only possibilities.  

By considering different coherence measures it can be shown that constant coherence corresponds to constant partition of support over symmetry subspaces. Our results extends these phenomena to mixed initial states and subchannels of higher rank than $1$. (The rank-1 case means that each subchannel is defined by a single Kraus operator).

\subsection{Purification}

Quantum trajectories that are initialised in a mixed state may converge to pure states at large times, which is known as \emph{purification}~\cite{maassen2006}. This is related to the mixedness of conditional states, $1-\Tr(\varrho_t^2)$, being a supermartingale under the stochastic dynamics determined by Kraus operators. Thus, it asymptotically converges to a random variable [by the martingale convergence theorem, cf. Eq.~\eqref{eq:conv}].
When that variable is $0$, the purification occurs, otherwise the mixedness of quantum trajectories is asymptotically constant. 

While this phenomenon is also related to martingale theory, it is separate from our results.
Namely, currently there is no quantum resource theory of mixedness (only unitary quantum channels would be free there). There exists a QRT of purity~\cite{Horodecki2003}, which measures {the quantum resource} of purity with the distance to the fully mixed state, which is the only free statein that theory.  However, this QRT does not consider stochastic free operations. 

Nevertheless, there is an indirect relation of our work to {the phenomenon of purification}, via dissipative freezing. If a mixed state is symmetric, it must remain symmetric even if it purifies, in which case such an asymptotic pure state must be supported on a single subspace. Quantum channels with strong symmetry belong to the set of strongly incoherent operations~\cite{streltsov2017}, so that the probability of finding conditional states in a given symmetry subspace is independent of their coherence before a last Kraus operator was applied. 
Therefore, when starting from a pure asymmetric state, quantum trajectories are asymptotically supported on pure symmetric states if and only if quantum trajectories for the corresponding dephased initial state  asymptotically purify.  (We note here that support on pure symmetric states is equivalent to support on a single symmetry subspace; and a dephased initial state is one where all coherences between symmetry subspaces have been removed.)
This shows that dissipative freezing can be considered not only for dynamics with strong symmetry, but more generally for strongly incoherent dynamics. 

 \subsection{Counterexample for {Theorem 4} of Ref.~\cite{schmolke2025}}

 While the phenomenon of purification in discrete dynamics is well understood~\cite{maassen2006,phdVanHorssen,Benoist2019,Benoist2024}, it is not  fully characterised in the continuous setting. This is an important issue, as it relates to the yet unknown general structure of asymptotic distributions in the stochastic quantum dynamics. Recently, such a characterisation was proposed in~\cite{schmolke2025}.  {In particular, {Theorem 4} of that work proposes conditions for the absence of purification, whose proof is conditional on a conjecture.  We present here a counterexample to that {Theorem 4}.}

First, note that discrete dynamics under a quantum channel $\Lambda$ can be directly extended to continuous-time setting with the master operator 
\beq
\mathcal{L}=\gamma(\Lambda-\mathcal{I}),
\label{eq:cts-time-op}
\eeq
 where $\mathcal{I}$ is the identity channel. This construction is achieved by applying $\Lambda$ at exponentially distributed waiting times with rate $\gamma$. Then discrete quantum trajectories are mapped to continuous-time quantum trajectories arising from counting measurements.  

Second, we note that for strictly incoherent operations (more specifically, dynamics with a strong symmetry), the purification occurs if only if the dissipative freezing does, and this in turn occur if and only if there is coherence vanishing. Kraus operators from Example II in the main text give rise to a strictly incoherent operation, and for initial states supported only within no-decay subspace, feature a strong symmetry.
Since coherence freezing occurs for pure states with $\alpha\neq 0$ or $\beta\neq0$, see the main text, the corresponding dephased states do not purify. This can also be seen directly from the orbits for mixed initial states [see Further information on Examples above {and Eqs.~(\ref{eq:sm:ex2_omega_1},\ref{eq:sm:ex2_omega_2})}]. 

 Finally, {Theorem 4} from~\cite{schmolke2025} proposes conditions for the absence of purification.
 Consider Example II and define subspaces denoted by {$Q_{g}$ and $Q_e$} which are spanned respectively by  $|g\rangle$ and $|e_1\rangle$, $|e_2\rangle$.  [These are the supports of two orthogonal steady states, cf. Eq.~\eqref{eq:sm:ex2_rho_ss}.]
Write $\mathbb{1}_{Q_j}$ for the orthogonal projection on subspace $Q_j$ (where $j=g,e$).
For the corresponding continuous-time dynamics of \eqref{eq:cts-time-op} the jump operators are $\sqrt{\gamma}K_m$,= with $m=1,...,4$. {As $Q_{g}$ and $Q_e$ are of different dimensions, the theorem states that  the absence of purification requires}
\begin{equation}
\mathbb{1}_{Q_j}K_m^\dagger K_m\mathbb{1}_{Q_j}\propto \mathbb{1}_{Q_j},
\label{eq:sm:schmolke}
\end{equation}
The continuous-time version of Example II does not purify {for a general initial state}, but
\begin{equation}
\mathbb{1}_{Q_e}K_1^\dagger K_1\mathbb{1}_{Q_e}= 2 |e_1\rangle\!\langle e_1|, \qquad\mathbb{1}_{Q_e}K_2^\dagger K_2\mathbb{1}_{Q_e}=2|e_2\rangle\!\langle e_2|\, .
\end{equation}
This clearly disproves {the condition in Eq.~\eqref{eq:sm:schmolke}} as $\mathbb{1}_{Q_e}=|e_1\rangle\!\langle e_1|+|e_2\rangle\!\langle e_2|$ [but does not seem to contradict Conjecture 3.1 from~\cite{schmolke2025}, as initial states supported only on $Q_g$ or $Q_e$ do purify, cf. Eqs.~(\ref{eq:sm:ex2_omega_3},\ref{eq:sm:ex2_omega_4})]. 
We emphasize that the condition for discrete dynamics not to purify in Theorem 1 from~\cite{maassen2006} remains satisfied (cf. Lemma 4 therein).

\end{document}